\documentclass[superscriptaddress,groupedaddress,nofootnoteinbib,11pt]{article}  

\usepackage{graphicx}  
\usepackage{dcolumn}   
\usepackage{bm}        
\usepackage{jcappub}

\def\be{\begin{equation}}
\def\ee{\end{equation}}
\def\ba{\begin{eqnarray}}
\def\ea{\end{eqnarray}}

\def\mpl{M_{\rm Pl}}

\def\d{\mathrm{d}}
\def\p{{\cal P}}

\def\L*{{\cal L}_*}
\def\L{\mathcal{L}}
\def\({\left(}
\def\){\right)}

\def\p{\partial}

\def\p{\partial}

\def\<{\langle}
\def\>{\rangle}

 \def\neq {\not\equiv}

\def\cs2{c_{s}^{2}}

 \def\p{\partial}

\begin{document}

\title{Low-Energy Effective Field Theory for Chromo-Natural Inflation}

\author{\hspace{-5pt}Emanuela Dimastrogiovanni$^a$,}
\author{Matteo Fasiello$^b$}
\author{and Andrew J. Tolley$^b$ \hspace{-40pt}}
\affiliation{$^a$School of Physics and Astronomy, University of Minnesota, Minneapolis, 55455, USA}
\affiliation{$^b$Department of Physics, Case Western Reserve University, 10900 Euclid Ave, Cleveland, OH 44106, USA}
\leftline{UMN-TH-3123/12}

\abstract{Chromo-natural inflation is a novel model of inflation which relies on the existence of non-abelian gauge fields interacting with an axion. In its simplest realization, an $SU(2)$ gauge field is assumed to begin inflation in a rotationally invariant VEV. The dynamics of the gauge fields significantly modifies the equations of motion for the axion, providing an additional damping term that supports slow-roll inflation, without the need to fine tune the axion decay constant. We demonstrate that in an appropriate slow-roll limit it is possible to integrate out the massive gauge field fluctuations whilst still maintaining the nontrivial modifications of the gauge field to the axion. In this slow-roll limit, chromo-natural inflation is exactly equivalent to a single scalar field effective theory with a non-minimal kinetic term, i.e. a $P(X,\chi)$ model. This occurs through a precise analogue of the {\it gelaton} mechanism, whereby heavy fields can have unsuppressed effects on the light field dynamics without contradicting decoupling. 
The additional damping effect of the gauge fields can be completely captured by the non-minimal kinetic term of the single scalar field effective theory. We utilize the single scalar field effective theory to infer the power spectrum and non-gaussianities in chromo-natural inflation and confirm that the mass squared of all the gauge field fluctuations is sufficiently large and positive that they completely decouple during inflation. These results confirm that chromo-natural inflation is a viable, stable and compelling model for the generation of inflationary perturbations. }

\maketitle


\section{Introduction}

Inflation is undoubtedly a very successful paradigm of modern cosmology as it naturally provides
an explanation for long standing issues such as the flatness and horizon problems. The simplest realizations of this idea are in excellent agreement with current CMB and LSS data. However, many inflationary models equipped with the above celebrated properties are not completely satisfactory when it comes to quantum corrections. For the most part in fact, scalar models rely on the fine tuning of the scalar  potential, a tuning that is generically not protected from large quantum corrections.  This is where symmetries of the action can come into play: if the flatness of the potential is tied to a (eventually broken) symmetry, the corrections to the potential will become manageable.  In this case we talk about the inflatiion model as being (technically) natural. \\

One such case consists of treating the scalar which drives inflation as a pseudo Nambu-Goldstone boson  endowed with a shift symmetry. The corrections to the potential will be proportional to the (small) symmetry breaking parameter, thus making the original flatness of the potential more natural. In the context of inflation, this idea was first implemented in  \cite{Freese:1990rb,Freese:2004un}, where an axion decay constant $f\sim M_{Pl}$ was required by observations. This fact turns out to be problematic and makes it hard to embed the model in string theory \cite{Banks:2003sx}. More recently, several models that relax this condition on the decay constant were put forward, notably in \cite{Kim:2004rp,Silverstein:2008sg,Anber:2009ua,Barnaby:2010vf,Adshead:2012kp}. In this manuscript we focus our attention on the specific model introduced in \cite{Adshead:2012kp}, \textit{chromo-natural} inflation. The appeal of this model is manifold: it successfully incorporates vector fields while at the same time preserving isotropy\footnote{Conversely, the fact that introducing vector fields naturally leads to anisotropy can be used precisely to mimic possible statistical anisotropy features in the CMB observations \cite{Yokoyama:2008xw, Watanabe:2009ct, Bartolo:2009pa,Dimopoulos:2009am, Gumrukcuoglu:2010yc}.} (see \cite{Galtsov:1991un} for early implementation of this idea, also \cite{Galtsov:2010dd, Maleknejad:2011jw, Maleknejad:2011sq, Galtsov:2011aa, Alexander:2011hz}); the way gauge fields couple to the axion is such
(preserving the axionic shift symmetry) that it is guaranteed that the higher order corrections to the gauge action will be small; the specific structure of the gauge coupling generates an additional damping effect for the axion, this provides further room to slow down for the evolution of the axion. Crucially this allows us to work with a natural model of inflation with $f \ll \mpl$, avoiding the fine-tuning in the original models of natural inflation. In our work we want to study this model in a specific scaling limit which, whilst simplifying the analysis, nevertheless captures all of the interesting features of the theory. For the benefit of the reader we quickly summarize the upshot of our analysis below.\\

The chromo-natural inflation model is specified by the action for an axion coupled to an $SU(2)$ gauge field:
\be
S_{\rm chromo}= \int \d^4 x \sqrt{-g} \left[ \frac{\mpl^2}{2} R - \frac{1}{4} F^a_{\mu \nu}F_a^{\mu \nu}- \frac{1}{2} (\partial \chi)^2 - \mu^4 \left( 1+ \cos\left( \frac{\chi}{f}\right) \right)+ \frac{\lambda}{8f }\chi \epsilon^{\mu \nu \rho \sigma} F^a_{\mu \nu}F^a_{\rho \sigma} \right] \, .
\ee
The main result of this work is the proof that when the mass of the gauge fields is sufficiently large for their fluctuations to decouple, chromo-natural inflation can be described by the following low energy effective field theory (EFT) for the axion:
\be
S_{\chi} = \int \d^4 x \sqrt{-g} \left[ \frac{\mpl^2}{2} R - \frac{1}{2} (\partial \chi)^2 +\frac{1}{4 \Lambda^4} \left(  \partial \chi \right)^4- \mu^4 \left( 1+ \cos\left( \frac{\chi}{f}\right) \right)  \right] +\dots \, ,
\ee
where $\dots$ signify $H^2/m_g^2$ suppressed corrections, $m_g$ is the mass of the gauge fields, and the scale $\Lambda$ is determined by $\Lambda^4 = 8 f^4 g^2 /\lambda^4$. The simplicity of this single field EFT  will allow us to easily obtain the power spectrum and bispectrum for chromo-natural inflation using known results for $P(X,\chi)$ models \cite{ArmendarizPicon:1999rj,Garriga:1999vw,Chen:2006nt}. The entire effect of the gauge fields is captured by the single modified kinetic term $ \left(  \partial \chi \right)^4/(4 \Lambda^4)$ which generates an additional damping, making it easier to satisfy the slow-roll condition even for a steep potential. 
The resulting EFT describes the full dynamics of gravity and the axion and should be distinguished from the EFT for the inflaton fluctuations (commonly denoted by $\pi$ in the literature) \cite{Cheung:2007st}. The above EFT is preferable to the one for $\pi$ in that it is not only valid around cosmological backgrounds; it also has a much wider regime of validity in energies (in particular it is IR complete). It is however straightforward to translate this into the language of $\pi$ as is well-known in the literature. We shall not do this here, preferring to work with the more elegant covariant form.  \\

This paper is organized as follows. In \textit{Section 2} we briefly review the chromo-natural inflation model and its main properties. In \textit{Section 3} we specialize our analysis to the \textit{slow roll hierarchy scaling limit} (SRHS) and show how the dynamics of the theory greatly simplifies in this limit, without losing any of its appeal. In \textit{Section 4} we show that the mass squared of the gauge fields, which are being integrated out, is positive and much larger than $H^2$; this guarantees a safe decoupling. In \textit{Section 5} we take on loop corrections showing that the low energy single-field effective field theory in technically natural. In \textit{Section 6} we elucidate how the dynamics of chromo-natural inflation can be understood as a clear  implementation of the so called \textit{gelaton} mechanism \cite{Tolley:2009fg}.  In \textit{Section 7} we use the machinery of $P(X,\chi)$ models of inflation to spell out the value of the observables extracted from our analysis, all well within experimental limits. The analysis of \textit{Section 3} is extended in \textit{Section 8} beyond the SRHS limit where a condition is identified for the effective single-field description to be valid. Further comments and conclusions
are to be found in \textit{Section 9}.\\

\section{Review of Chromo-Natural Inflation}

\noindent In this section we will briefly review chromo-natural inflation at the background level. The initial Lagrangian is 
\be
S_{\rm chromo}=\int \d^{4}x\sqrt{-g}\left[\frac{\mpl}{2}R-\frac{1}{4}F_{\mu\nu}^{a}F_{a}^{\mu\nu}-\frac{1}{2}\left(\partial\chi\right)^{2}-\mu^{4}\left(1+\cos\left(\frac{\chi}{f}\right)\right)+\frac{\lambda}{8f}\chi\epsilon^{\mu\nu\rho\sigma}F_{\mu\nu}^{a}F_{\rho\sigma}^{a}\right]
\ee
where $F_{\mu\nu}^{a}=\partial_{\mu}A^{a}_{\nu}-\partial_{\nu}A^{a}_{\mu}-g f^{abc}A_{\mu}^{b}A_{\nu}^{c}$ and $\chi$ is the axion field ($f^{abc}$ is the 3 dimensional Levi-Civita symbol normalized with $f^{123}=1$). The background metric is the standard inflationary FRW, $ds^{2}=-dt^{2}+a^{2}(t)\delta_{ij}dx^{i}dx^{j}$. We will work with the convention $\epsilon^{0123}=\frac{1}{\sqrt{-g}}$.\\

The $SU(2)$ gauge freedom allows for an isotropic configuration for the spatial triplet and for a zero temporal component of the gauge field:
\be
A_{0}^{a}=0,
\ee
\be
\label{bA1}
A_{i}^{a}=\delta^{a}_{i}a(t)\psi(t).
\ee
Given this, all  of the background equations of motion follow from the minisuperspace action
\ba
S_{\rm chromo} &=& \int \d^4 x N a^3 \left[ - 3 \mpl^2 \frac{\dot a^2}{N^2 a^2} + \frac{1}{2 N^2} {\dot{\chi}}^2 -\mu^4 \left(1+\cos \left( \frac{\chi}{f}\right) \right) \right.\\
&& \left. +\frac{3}{2N^2a^2} \left(\frac{\partial (\psi a)}{\partial t} \right)^2-\frac{3}{2}g^2 \psi^4 + \frac{g \lambda}{f N} \dot \chi \psi^3 \right]
\ea
(where $N$ can be set to unity after variation).
The Friedman equation (varying with respect to $N$) reads
\be\label{ee1}
H^2 =\frac{1}{3 \mpl^2} \left[\frac{3}{2a^2}\left(\partial_{t}\left(a\psi\right)\right)^2+\frac{3}{2}g^{2}\psi^{4}+\frac{\dot{\chi}^{2}}{2}+\mu^{4}\left(1+\cos\left(\chi/f\right)\right) \right].
\ee
One important condition for inflation to occur in this model is that the potential of the axion dominates the total energy density:
\be\label{sl0}
3H^{2}\mpl^2 \simeq \mu^{4}\left(1+\cos\left(\chi/f\right)\right).
\ee
The equations of motion for the axion and the gauge field are: 
\be\label{fe1}
\ddot{\chi}+3H\dot{\chi}-\frac{\mu^{4}}{f}\sin\left(\frac{\chi}{f}\right)=-\frac{3g\lambda}{f}\psi^2\left(\dot{\psi}+H\psi\right),
\ee
\be\label{fe2}
\ddot{\psi}+3H\dot{\psi}+\left(\dot{H}+2H^{2}\right)\psi+2g^2\psi^3=\frac{g\lambda}{f}\psi^{2}\dot{\chi}.
\ee
In the slow-roll regime, inflationary solutions have been found for the system. Indeed, by setting $\ddot{\chi}\simeq 0$, $\ddot{\psi}\simeq 0$ and $\dot{H}\simeq 0$ in Eqs.~(\ref{fe1})-(\ref{fe2}), one can solve for $\dot{\psi}$ and $\dot{\chi}$ \cite{Adshead:2012kp}; the result for the gauge field time derivative is
\be
\dot{\psi}=-\frac{H\psi\left(2f^{2}H^{2}+2g^{2}f^{2}\psi^{2}+g^{2}\lambda^{2}\psi^{4}\right)}{3f^{2}H^{2}+g^{2}\lambda^{2}\psi^{4}}+\frac{g\lambda\psi^{2}\mu^{4}\sin(\chi/f)}{3\left(3f^{2}H^{2}+g^{2}\lambda^{2}\psi^{4}\right)},
\ee
and, if we assume $3f^{2}H^{2}\ll g^{2}\lambda^{2}\psi^{4}$ and $\lambda \psi /f \gg 1$, we are left with:
\be
\label{pp1}
H\dot{\psi}\simeq -H^{2}\psi\left(1-\frac{\mu^{4}\sin(\chi/f)}{3g\lambda H\psi^{3}}\right).
\ee
The latter can be rewritten as an equation for the gauge field, $H\dot{\psi}+\frac{\partial V_{eff}}{\partial\psi}=0$, with an effective potential
\be
V_{eff}\equiv \frac{H^{2}\psi^{2}}{2}+\frac{H\mu^{4}\sin(\chi/f)}{3g\lambda\psi},
\ee
which happens to be minimized by
\ba\label{gfm}
\psi_{min}\simeq \left(\frac{\mu^{4}\sin(\chi/f)}{3g\lambda H}\right)^{1/3}.
\ea
This solution provides an effective flat potential for the axion or, putting it in another way, the solution increases the effective damping for the motion of the axion. This may be made most transparent by looking at the effective slow-roll equation for the axion which, from \ref{fe1}, is
\be
3H \left( 1+ \frac{\lambda^2 \psi^2}{2f^2}\right)\dot{\chi}-\frac{\mu^{4}}{f}\sin\left(\frac{\chi}{f}\right) \simeq 0 \, .
\ee
This equation makes explicit the fact that the gauge field creates an additional damping term for the axion equation of motion.\\

Eq.~(\ref{sl0}) holds if the kinetic energies of $\chi$ and $\psi$, along with the gauge field potential, are assumed to only contribute in a negligible fraction to the total energy density of the Universe. We can therefore define some ``small parameters'' for the system:
\begin{eqnarray}\label{sl2}
&&\epsilon_{\chi}\equiv \frac{\dot{\chi}^{2}}{2H^{2}\mpl^2},\\ \label{sl3}
&&\epsilon_{\psi}\equiv \frac{\dot{\psi}^{2}}{H^{2}\mpl^2},\\\label{sl4}
&&\epsilon_{1}\equiv \frac{\psi^{2}}{\mpl^2},\\\label{sl5}
&&\epsilon_{2}\equiv \frac{g^{2}\psi^{4}}{H^{2}\mpl^2}.
\end{eqnarray}
From the Einstein equations
\be\label{ee2}
\dot{H}=-\frac{1}{M_{Pl}^{2}}\left[\frac{1}{a^2}\left(\p_{t}\left(a\psi\right)\right)^{2}+g^{2}\psi^{4}+\frac{\dot{\chi}^{2}}{2}\right],
\ee
we can see how the slow-roll parameter $\epsilon\equiv-\dot{H}/H^{2}$ arises from a sum of all the small parameters (\ref{sl2}) through (\ref{sl5}). The regime in which the gauge field settles at its minimum (\ref{gfm}), however, implies a hierarchy between the fields and their time derivatives, for instance it is easy to see from (\ref{pp1}) that we automatically have $\epsilon_{\psi} \ll \epsilon_1$. Therefore, from now on, the relevant slow-roll parameters  will be: 
\be
\epsilon\simeq \epsilon_{1}+\epsilon_{2}+\epsilon_{\chi}.
\ee

\section{Slow-roll Hierarchy Scaling Limit}

The slow-roll hierarchy we intend to work in,  $\epsilon_1 \ll \epsilon_2 \ll 1$, is equivalent to requiring that the mass of the gauge fields $m_g= \sqrt 2 g \psi$ (we shall prove below that this is indeed the scale that sets the gauge field mass in \textit{Section} \ref{gaugemass}) is much greater than the Hubble scale $H$ since
\be
\frac{m_g^2}{2 H^2} = \frac{\epsilon_2}{\epsilon_1} \gg  1 \, .
\ee
A simple way to capture this hierarchy is to perform the following redefinition of the fields
\be
\psi = \frac{1}{\sqrt{g}} \, \hat \psi \, , \quad     A_{\mu}^a =  \frac{1}{\sqrt{g}}  \hat  A_{\mu}^a \  \, , \quad
 \lambda = \sqrt{g} \, \hat \lambda \, ,
\ee
so that
\be
\epsilon_1= \frac{\hat \psi^2}{g \mpl^2} , \, \quad  
\epsilon_2 = \frac{\hat \psi^4 }{H^2 \mpl^2} \, .
\ee
We now see that the slow-roll hierarchy $ \epsilon_2 \gg \epsilon_1$  can be captured by formally taking $g \rightarrow \infty$ keeping $\hat{\psi}$ (and $\hat{\lambda}$, as we shall see below) fixed. In other words, we shall use the expansion parameter ${1}/{g}$ to keep track of the slow-roll dependence $\epsilon_2/\epsilon_1$. 
This is what we are going to refer to as the slow-roll hierarchy scaling limit (SRHS limit) in what follows. It is important to stress that taking this limit does not imply that the gauge field coupling $g$ is larger than unity, the mathematical scaling limit $g \rightarrow \infty$ is rather encoding the physical requirement that
\be
g \gg  \frac{\hat \psi^2}{\epsilon_1  \mpl^2} \, ,
\ee
which can be satisfied even when $g \ll 1$ due to the largeness of $\mpl$. This type of scaling limit can also be performed for the gelaton model \cite{Tolley:2009fg} and in \textit{Section} \ref{gelaton} we shall see that chromo-natural inflation is an explicit example of the gelaton mechanism. \\

Before we proceed, we must check that in this limit the nontrivial effects of the gauge field observed in \cite{Adshead:2012kp} are not lost. To this aim, we simply take the SRHS limit of the background equations of motion: 
\be
\ddot{\chi}+ 3 H \dot{\chi}-\frac{\mu^4}{f} \sin \left( \frac{\chi}{f}\right)=-3 \frac{\hat \lambda}{f} \hat \psi^2 \left( \dot {\hat \psi}+H \hat \psi \right) \, , \quad  2  \hat \psi^3 = \frac{\hat \lambda }{f}  \hat \psi^2 \dot \chi \, .
\ee
The second equation simplifies in this limit and we immediately see that
\be
\label{slow1}
\hat \psi =  \frac{\hat \lambda }{2 f}  \dot \chi \, .
\ee
Substituting in the first equation and taking the slow-roll limit by neglecting $\ddot{\chi}$ we have
\be
\label{slow2}
3 \(1  + \frac{\hat \lambda^2}{2 f^2} \hat \psi^2  \)H \dot \chi \simeq \frac{\mu^4}{f} \sin  \left( \frac{\chi}{f}\right) \, .
\ee
By taking the limit ${\hat \lambda^2  \hat \psi^2 }/{f^2}\gg 1$ we see that the gauge field enhances the Hubble damping term. This is the central novel effect of chromo-natural inflation. The gauge field VEV damps the evolution of the axion making it easier to satisfy the slow-roll conditions even on a steep potential. This is precisely the same effect used in DBI inflation \cite{Alishahiha:2004eh}, k-inflation \cite{ArmendarizPicon:1999rj,Garriga:1999vw} and Galileon inflation/G-inflation models \cite{Burrage:2010cu}\cite{Kobayashi:2010cm}.
 
Finally combining the previous two equations we recover that 
\be
\hat \psi \simeq \left( \frac{\mu^4  \sin  \left( {\chi}/{f}\right)}{3 \hat \lambda H}\right)^{1/3} \, ,
\ee
which is the same as the equation used in \cite{Adshead:2012kp}, namely \ref{gfm}.

Thus we see that the SRHS limit still captures all of the essential features of chromo-natural inflation. For future convenience, we note that we can also express the equation (\ref{slow2}) using (\ref{slow1}) in the form:
\be
\label{slow3}
3 \(1  + \frac{\hat \lambda^4}{8 f^4} \dot{\chi}^2  \)H \dot \chi \simeq \frac{\mu^4}{f} \sin  \left( \frac{\chi}{f}\right) \, .
\ee
The fact that this is a single effective equation for $\chi$ will be extremely significant in what follows. Indeed, the quickest (although least rigorous) way to our main result would be to write down a scalar field action whose slow-roll equations reproduce this equation. Rather than doing this, we shall now show that we can obtain this result by consistently integrating out the gauge fields. In \textit{Section} \ref{beyond} however, we will use the quick route to make a conjecture for an improved version of the single field EFT which is valid beyond the regime of validity of the SRHS limit. 
\\

\subsection{Deriving the single field EFT}

We now extend the SRHS limit to the full action including perturbations. We begin with the full action:
\be
S_{\rm chromo}= \int \d^4 x \sqrt{-g} \left[ \frac{\mpl^2}{2} R - \frac{1}{4} F^a_{\mu \nu}F_a^{\mu \nu}- \frac{1}{2} (\partial \chi)^2 - \mu^4 \left( 1+ \cos\left( \frac{\chi}{f}\right) \right)+ \frac{\lambda}{8f }\chi \epsilon^{\mu \nu \rho \sigma} F^a_{\mu \nu}F^a_{\rho \sigma} \right]\, .
\ee
In the limit $g \rightarrow \infty$, keeping $\hat \lambda$ and $\hat A_{\mu}^a$ fixed, we find
\ba
S_{\rm SRHS}&=& \int \d^4 x \sqrt{-g} \left[ \frac{\mpl^2}{2} R - \frac{1}{4} f^{abc}\hat A_{\mu}^b \hat A_{\nu}^c f^{ade} \hat A^{\mu}_d \hat A^{\nu}_e- \frac{1}{2} (\partial \chi)^2 - \mu^4 \left( 1+ \cos\left( \frac{\chi}{f}\right) \right)  \right. \nonumber \\
&&\left. -\frac{\hat \lambda}{2f }\chi \epsilon^{\mu \nu \rho \sigma} \partial_{\mu} \hat A_{\nu}^a f^{abc} \hat A_{\rho}^b  \hat A_{\sigma}^c \right] \, .
\ea
The scaling of $\lambda=\sqrt{g} \hat \lambda$ was chosen precisely so that the axion coupling to the gauge field survives in the limit. Note that, because of the maximal antisymmetry of the Levi-Cevita symbol, this coupling includes at least one derivative. 
After an integration by parts, we have:
\ba
S_{\rm SRHS}&=& \int \d^4 x \sqrt{-g} \left[ \frac{\mpl^2}{2} R - \frac{1}{4}  (\hat A_{\mu}^a \hat A^{\mu}_a)^2+\frac{1}{4} (\hat A_{\mu}^a \hat A_{\nu}^a) (\hat A^{\mu}_b \hat A^{\nu}_b)  - \frac{1}{2} (\partial \chi)^2 - \mu^4 \left( 1+ \cos\left( \frac{\chi}{f}\right) \right)  \right.  \nonumber \\
&& \left. + \frac{\hat \lambda}{6f } \partial_{\mu}\chi \epsilon^{\mu \nu \rho \sigma} f^{abc} \hat A_{\nu}^a \hat A_{\rho}^b \hat A_{\sigma}^c\right] \, .
\ea
We see that in this limit the gauge field only enters algebraically, and, in addition, it still preserves the full $SU(2)$ gauge symmetry. The original realization of the symmetry in infinitesimal form
\be
\delta_{\theta} A_{\mu}^a = \frac{1}{g}\partial_{\mu} \theta^a -  f^{abc} A_{\mu}^b \theta^c \, ,
\ee
is replaced with the $SU(2)$ symmetry transformation
\be
\delta_{\theta} \hat A_{\mu}^a = - f^{abc}  \hat A_{\mu}^b  \theta^c \, ,
\ee
which corresponds to the statement that in this limit the gauge field transforms normally (rather than as a connection) in the adjoint representation of the gauge group. We can for example use this gauge freedom to ensure that the $3 \times 3 $ matrix whose components are $A_i ^ a$, where $i=x,y,z$, is symmetric. We shall utilize this fact to great effect in the next section. \\
For now, we note that, since the action $S_{\rm SRHS}$ is actually only quadratic in $A_0^a$, we may straightforwardly integrate it out. The remaining equation for $A_i^a$ is difficult to solve directly, however we can obtain the answer with a trick: since the action is algebraic in $A_{\mu}^a$, it is sufficient to solve it at a single space-time point $x_P$ in a locally inertial rest frame. In other words, we may take the metric near the point $x_P$ to be of the form $g_{\mu \nu}  = \eta_{\mu \nu} + {\cal O}((x-x_P)^2)$. Special relativity tells us that we may also express the time-like space-time vector $\partial_{\mu} \chi$ at the point $x_P$ as a local Lorentz boost of a vector pointing entirely in the time direction with the same Lorentz invariant norm:
\be
\partial_{\mu} \chi(x_P) = \Lambda_{\mu}{}^{\nu}(x_P) \delta_{\nu 0} \sqrt{- (\partial \chi (x_P))^2}\, ,
\ee
where $\Lambda_{a}{}^{A}(x_P)\Lambda_{b}{}^{B}(x_P)\eta_{AB}=\eta_{ab}$. Lorentz boosting the gauge field in the same way:
\be
\hat A_{\mu}^a(x_P)  =  \Lambda_{\mu}{}^{\nu}(x_P) \, \tilde A_{\nu}^a(x_P)  \, ,
\ee
we may now use the local Lorentz invariance of the action to infer that $ \Lambda_{\mu}{}^{\nu}(x_P)$ drops out so that the relevant part of the Lagrangian at $x_P$ is: 
\ba
{\cal L}_A(x_P)& =&  - \frac{1}{4}  (\tilde A_{\mu}^a \tilde A^{\mu}_a)^2+\frac{1}{4} (\tilde A_{\mu}^a \tilde A_{\nu}^a) (\tilde A^{\mu}_b \tilde A^{\nu}_b)  + \frac{\hat \lambda}{6f } \delta_{\nu 0} \sqrt{- (\partial \chi (x_P))^2} \epsilon^{\mu \nu \rho \sigma} f^{abc} \tilde A_{\nu}^a \tilde A_{\rho}^b \tilde A_{\sigma}^c \nonumber \\
 &=& \frac{1}{2} (\tilde A_0^a)^2(\tilde A_i^b)^2-\frac{1}{2} (\tilde A_0^a \tilde A_i^a) (\tilde A_0^b \tilde A_i^b) \nonumber 
 \\ &&- \frac{1}{4}  (\tilde A_{i}^a \tilde A^{i}_a)^2+\frac{1}{4} (\tilde A_{i}^a \tilde A_{j}^a) (\tilde A^{i}_b \tilde A^{j}_b)  + \frac{\hat \lambda}{6f } \sqrt{- (\partial \chi (x_P))^2} \epsilon^{ij k} f^{abc} \tilde A_{i}^a \tilde A_{j}^b \tilde A_{k}^c \, .
\ea
Recognizing that the action is quadratic in $\tilde A_0^a$, varying with respect to it and assuming $\tilde A_i^a$ is nonzero, we find:
\be
\tilde A_0^a(x_P) =0 \, .
\ee
 We can then substitute it back into the action:
\be
{\cal L}_A(x_P)=- \frac{1}{4}  (\tilde A_{i}^a \tilde A^{i}_a)^2+\frac{1}{4} (\tilde A_{i}^a \tilde A_{j}^a) (\tilde A^{i}_b \tilde A^{j}_b)  + \frac{\hat \lambda}{6f } \sqrt{- (\partial \chi (x_P))^2} \epsilon^{ij k} f^{abc} \tilde A_{i}^a \tilde A_{j}^b \tilde A_{k}^c \ .
\ee
Varying with this with respect to $\tilde A_i^a$ we find a cubic equation which allows for a number of distinct branches of solutions. However, the branch which includes the solutions utilized in the background solution for chromo-natural inflation is simple to identify and we find:
\be
\tilde A_i^a(x_P) = \delta_i^a \frac{\hat \lambda}{2 f} \sqrt{- (\partial \chi (x_P))^2} \, .
\ee
The validity of this solution can be easily checked by comparing with the background equation (\ref{bA1}). In the next section we shall show that this solution is stable under perturbations. \\

Finally, restricting ourselves to this branch of solutions and substituting back into the Lagrangian, we find
\be
{\cal L}_A(x_P) =\frac{1}{4 \Lambda^4} \left(  \partial \chi(x_P) \right)^4\, .
\ee
Although we evaluated this in a locally inertial frame, it is now straightforward to covariantize this expression to determine the general result. Thus, the action for the gauge field is equivalent to a non-minimal kinetic term for the axion.  \\

Putting this all together we finally obtain the single field effective action:
\be
S_{\chi} =S_{\rm SRHS}= \int \d^4 x \sqrt{-g} \left[ \frac{\mpl^2}{2} R - \frac{1}{2} (\partial \chi)^2 +\frac{1}{4 \Lambda^4} \left(  \partial \chi \right)^4- \mu^4 \left( 1+ \cos\left( \frac{\chi}{f}\right) \right)  \right]\, ,
\ee
where $\Lambda^4 = 8 f^4/\hat \lambda^4=8 f^4 g^2 /\lambda^4 $. This is the central observation of this paper, chromo-natural inflation is, in the SRHS limit, precisely equivalent to a single scalar field effective theory with a classic Goldstone-like $P(X,\chi)$ Lagrangian where 
\be
P(X,\chi)= X+\frac{1}{\Lambda^4}X^2- \mu^4  \left( 1+ \cos\left( \frac{\chi}{f}\right) \right) \, ,
\ee
and $X = - (1/2)(\partial \chi)^2$. Since we have already show that the SRHS limit still captures the essential features of chromo-natural inflation at the level of the background, it is clear that the above scalar field Lagrangian must capture the same. In particular, the extra friction term that arises in the effective slow-roll equation for the axion in this model comes from the existence of the non-minimal kinetic term.  \\

In fact, writing down the background equation of motion for the axion in this effective theory we have simply:
\be
\ddot{\chi} \left(1+ \frac{3 \dot \chi^2}{\Lambda^4} \right) + 3 H \dot{\chi} \left(1+ \frac{\dot \chi ^2}{\Lambda^4} \right) -\frac{\mu^4}{f} \sin\left( \frac{\chi}{f}\right) =  0 \, .
\ee
Working in the slow-roll limit in which we assume we can neglect $\ddot{\chi}$, we find:
\be
3 H \dot{\chi} \left(1+ \frac{\dot \chi ^2}{\Lambda^4} \right) -\frac{\mu^4}{f} \sin\left( \frac{\chi}{f}\right) =  0 \, .
\ee
This is identical to equation (\ref{slow3}), which was obtained by combining (\ref{slow2}) with (\ref{slow1}). This confirms that the entire effect of the extra damping term arising in chromo-natural inflation can be completely captured by a simple modified kinetic term. In other words, the novel features of chromo-natural inflation that distinguish it from normal axion inflation, are those that arise when $\frac{1}{4 \Lambda^4} \left(  \partial \chi \right)^4$ dominates over $ \frac{1}{2} (\partial \chi)^2 $. However, the above analysis tells us that this works not just for the background evolution but also for the perturbations. Since we have derived the single field effective theory as a consistent scaling limit, we can consistently use the single field low energy effective field theory (EFT) to compute perturbations safe in the knowledge that the corrections from the gauge field will be suppressed by positive powers of ${H}/{m_g}$.

\section{Mass of Gauge Field Fluctuations and Stability}

\label{gaugemass}

In the previous section we integrated out the gauge fields at tree level in the SRHS limit (we shall consider loops in the next section) to derive the low energy effective theory for the axion. However, an important check on the consistency of this process is that the mass squared of all the gauge field fluctuations are indeed large and, crucially, that they are also all positive\footnote{Scaling limits have the unfortunate feature that, although they nicely capture the decoupling of heavy degrees of freedom, they cannot distinguish between a large positive mass squared and large negative mass squared. Physically a large negative mass would signal a disastrous instability for the system. It thus behoves us to demonstrate, in addition to the existence of a well-defined scaling limit, that indeed all the masses are the heavy fields are positive definite.}. This is necessary for the stability of the theory and thus consistency of the low energy EFT.  

Returning to the original action before we take the SRHS limit, the gauge field contributions are given by:
\be
S_{\rm gauge}= \int \d^4 x \sqrt{-g} \left[ - \frac{1}{4} F^a_{\mu \nu}F_a^{\mu \nu}+ \frac{\lambda}{8f }\chi \epsilon^{\mu \nu \rho \sigma} F^a_{\mu \nu}F^a_{\rho \sigma} \right] \, .
\ee
We see that there are two contributions to the mass term, one directly from the Yang-Mills action, and a second which arises from the axion coupling after we have integrated by parts and placed the time-derivative on $\chi$. In determining the mass of the gauge fields, it is sufficient to perturb only the gauge fields themselves and neglect the kinetic and mass mixing with the axion $\chi$ and with gravity $g_{\mu\nu}$. The reason for this is that, as will be justified retroactively, the gauge field fluctuations will be far more massive than those of the axion or graviton.  Given the existence of the large mass term, any cross-coupling can be removed by a local field redefinition of $B_{\mu}^a$. To give an example, given a schematic action for a light field $\chi$ and a heavy field $B$
\be
S= \frac{1}{2}B [\Box-m^2] B - \lambda B \chi + \frac{1}{2} \chi \Box \chi \dots  \, ,
\ee
we can diagonalize via $B = B'+  \lambda [\Box-m^2] ^{-1} \chi$ to give
\be
S= \frac{1}{2}B'  [\Box-m^2]  B' - \frac{\lambda^2}{2} {\chi} \ [\Box-m^2] ^{-1} {\chi}  + \frac{1}{2} \chi \Box \chi+\dots  \, .
\ee
Although these formal steps may always be performed, it is only when $B$ is far more massive than both $\chi$ and  the energy scales being probed (i.e. $H$), that it is possible to expand the operator $[\Box-m^2] ^{-1} $ in terms of local operators
\be
 [\Box-m^2] ^{-1} = -\frac{1}{m^2}- \frac{1}{m^4}\Box + \dots
\ee
In this case it is safe to interpret $m$ as the mass of $B$ without the need to diagonalize. The same is not true for the light field which picks up a mass of the form $m^2_{\chi} \sim - \lambda^2/m^2$.
In summary, assuming from the outset a hierarchy in the masses, it is sufficient to calculate the mass of the gauge field fluctuations assuming we are working on a fixed FRW background with $\chi$ only a function of time. Once the masses are known, the existence of the hierarchy is then justified after the fact. \\

In order to compute the mass it is helpful to integrate by parts from the outset to put the action in a form in which it is manifestly invariant under the axion shift symmetry:
\ba
S_{\rm gauge}&=& \int \d^4 x \sqrt{-g} \left[ - \frac{1}{4} F^a_{\mu \nu}F_a^{\mu \nu}- \frac{ \lambda}{2f } \partial_{\mu} \chi \epsilon^{\mu \nu \rho \sigma} A^a_{\nu}\partial_{\rho} A^a_{ \sigma} +\frac{g \lambda}{6f } \partial_{\mu} \chi \epsilon^{\mu \nu \rho \sigma} f^{abc }A^a_{\nu}A^b_{\rho} A^c_{ \sigma}  \right]  \nonumber \\
&=& \int \d^4 x \sqrt{-g} \left[ - \frac{1}{4} F^a_{\mu \nu}F_a^{\mu \nu}- \frac{ \lambda}{2f } \dot \chi \epsilon^{ijk} A^a_{i}\partial_{j} A^a_{ k} +\frac{g \lambda}{6f } \dot  \chi \epsilon^{ijk} f^{abc }A^a_{i}A^b_{j} A^c_{k}  \right] , \label{gaugeaction}
\ea
(here $\epsilon^{123}=1/\sqrt{-g}$). Because this is a gauge theory, the $A_0^a$ are not physical degrees of freedom, but rather act, in the phase-space formulation, as Lagrange multipliers for the non-Abelian generalization of the Gauss law constraint. To make it more transparent, we rewrite this action in phase-space form by defining the momentum conjugate
\be
\pi^i_a = F^{0 i }_a \, ,
\ee
so that the phase-space action is given by
\be
S_{\rm gauge}=\int \d^4 x \sqrt{-g} \left[ \pi^i_a \frac{\partial}{\partial t} A_i^a+ A_0^a D_i \pi^i_a- {\cal H }_{\rm gauge} \right] \, ,
\ee
where the Hamiltonian density is given by
\be
{\cal H }_{\rm gauge} = \frac{1}{2} \pi^i_a \pi_i^a+ \frac{1}{4} F^a_{i j}F_a^{i j}+  \frac{ \lambda}{2f } \dot \chi \epsilon^{ijk} A^a_{i}\partial_{j} A^a_{ k} -  \frac{g \lambda}{6f } \dot  \chi \epsilon^{ijk} f^{abc }A^a_{i}A^b_{j} A^c_{k}  \, .
\ee
Each gauge field carries two physical degrees of freedom, making a total of 6. These physical degrees of freedom can be captured by 6 of the spatial components of the gauge field $A_i^a$ with the remaining 3 components of $A_i^a$ being removable by an $SU(2)$ gauge transformation. We shall see below that there is a natural choice for the gauge fixing which corresponds to choosing the symmetric gauge condition
\be
A_{ia} =A_{ai} \, , \label{symmetric}
\ee
which mixes gauge indices and space indices.
The Gauss law constraint $D_i \pi^i_a=\partial_i \pi^i_a - g f^{abc} A_{i}^b \pi^i_c=0$ is used to remove the 3 corresponding conjugate momenta as is usual in Yang-Mills theories.

To determine the mass of the gauge fields we perturb around the cosmological background and decompose 
\be
A_{i}^a=\bar A_{i}^a + B_{i}^a \, , \quad \pi^{i}_a=\bar \pi^{i}_a + P^{i}_a \, , \quad A_0^a = 0 + \lambda_0^a
\ee
where $ \bar A_{\mu}^a$ is the background solution, $B_{\mu}^a$ denotes the gauge field fluctuations and with $\bar \pi^{i}_a $ and $P^{i}_a$ their associated conjugate momenta. To infer the mass for the gauge field fluctuations it is sufficient to expand the action $S_{\rm gauge}$ to second order in $B_{\mu}^a$, neglecting spatial derivatives of $B_{\mu}^a$ since these contribute only to the kinetic term and the gradient energy\footnote{We are implicitly neglecting the contributions to the mass of order $H$ and $\dot{H}$ coming from derivatives of the scale factor which would arise if we were to canonically normalize the gauge fields fluctuations precisely because we are working in the SRHS limit $m_g \gg H$.}. \\

\noindent Linearizing the Gauss-law constraint we have
\be
\partial_i P^i_a -g f^{abc}\bar A^b_i P^i_c-g f^{abc}  B^b_i \bar \pi^i_c =0 \, .
\ee
This equation places a constraint on $P^i_c$ which in the SRHS limit (or long wavelength limit, which is sufficient to calculate the mass terms) gives:
\be
 f^{aic} a \psi \, P^i_c+ f^{abi}  B^b_i \frac{\partial}{\partial t}(a \psi) \approx 0 \, .
\ee
In the SRHS limit the gauge freedom on $B_{i}^a$ amounts to
\be
\delta B_{i}^a = - f^{abc}  a \hat \psi \delta_i^b  \theta^c = - f^{a i c} \theta^c (a \hat \psi) \, .
\ee
In other words, under an infinitesimal gauge transformation, the $3 \times 3$ matrix $B$ whose components are $B_{i}^a $ transforms only via an antisymmetric matrix since $f^{aic}=-f^{iac}$. Given this fact, we may always choose a gauge for which $B$ is symmetric. 
If we choose to work in the gauge for which $B_{ia} =B_{ai}$, then the Gauss-law constraint implies:
\be
P^{ia} \approx P^{ai} \, .
\ee
That is, it is the symmetric components of the momentum that are conjugate to the symmetric gauge field. Having solved for the constraint, we may substitute this information back into the action and the Lagrange multiplier $\lambda_0^a$ drops out. This amounts to little more than remembering that $B_{ia}$ and $P^{ia}$ are symmetric. \\

Given these considerations, it is clear that the mass term comes entirely from the spatial components of the gauge field $B_i^a$ which come from the non-derivative terms in $\cal H_{\rm gauge}$ 
\be
{\cal H}_{\rm gauge} =  \frac{1}{2a^2} g^2 \psi ^2  \left[ 2 (B_i^i)^2- B_i ^j B_j^i +(B_{i}^a)^2 \right]  - \frac{g \lambda \dot{\chi}}{2 f a^2} \psi  \left[ (B_i^i)^2- B_i ^j B_j^i\right] + \dots\, .
\ee
Utilizing the fact that in the SRHS limit $\dot{\chi}=2 g \psi  f/\lambda$, and the symmetric properties of $B_{ia}$ in our chosen gauge gives
\be
{\cal H}_{\rm gauge} =  \frac{1}{2a^2} 2 g^2 \psi^2  \sum_{a,i} \left(B_{i}^a \right)^2 +\dots \, .
\ee
We see that all of the 6 remaining physical components of the gauge field have positive mass squared terms with the same mass $m_g= \sqrt 2 g \psi$.
For reference, the gauge action to quadratic order is then:
\begin{eqnarray}
 S^{(2)}_{\rm gauge} = \int \d^4 x &\sqrt{-g}&\Big[ P^i_a \frac{\partial}{\partial t} B_i^a-\frac{1}{2} P_i^a P^i_a-\frac{1}{4 a^4} (\partial_i B_j^a-\partial_j B_i^a)^2- \frac{1}{2a^2} 2 g^2 \psi^2   \left(B_{i}^a \right)^2    -2g\psi \epsilon^{ijk} B^a_i \p_j B^a_k \nonumber \Big]\\
= \int \d^4 x &\sqrt{-g}& \Big[ \frac{1}{2a^2} \dot B_i^a \dot B^i_a-\frac{1}{4a^4} (\partial_i B_j^a-\partial_j B_i^a)^2- \frac{1}{2a^2} 2 g^2 \psi^2   \left(B_{i}^a \right)^2 -2g\psi \epsilon^{ijk} B^a_i \p_j B^a_k 
\Big]  \, ,
\ea
along with the condition that $B_{ia}=B_{ai}$. Although in this representation the gradient energy is not diagonalized so there there is still a coupling between the different components of $B_{ia}$, this coupling is irrelevant for $m_g \gg k/a$ and so does not affect the determination of the masses.  \\

We have thus shown that all of the 6 physical degrees of freedom of the $SU(2)$ gauge fields have the same positive mass $m_g=\sqrt{2} g \psi$ which is parametrically larger than $H$ in the SRHS limit, retroactively justifying the neglect of the couplings of the gauge fields to the axion and metric fluctuations and confirming that the model is indeed stable.

\subsection*{Gradient Instability}

We now move on to analyze the potential existence of a gradient instability. The starting point is that of Eq.(\ref{gaugeaction}) plus the kinetic term for the axion. With the large mometum region in mind $(k^2/a^2\gg m_g^2 \gg H^2)$, we focus on the terms in the action equipped with two derivatives (both space and time). In this regime it is easier to analyze the system by adding to the action the Lorenz gauge fixing term appropriate to a Lorenz gauge Fadeev-Popov quantization rather then implementing the symmetric gauge condition of Eq.~(\ref{symmetric}). \\

The two-derivative terms for the gauge field and axion are
\ba
S_{\rm two-deriv} = \int \d^4 x \sqrt{-g} \left[  \, -\frac{1}{2}(\p_{\mu}\chi)^2 -\frac{1}{4} (\p_{\mu} A^{a}_{\nu}-\p_{\nu} A^{a}_{\mu})^2 -\frac{ \lambda}{2 f}\partial _{\mu}\chi \epsilon^{\mu \nu \rho \sigma} A^{a}_{\nu}\p_{\rho}A^{a}_{\sigma} -\frac{1}{2}(\p_{\mu}B^{\mu\,a})^2  \right]\nonumber
\ea
where the last term is the gauge fixing term associated with the Lorenz gauge $\partial_{\mu} B^{\mu\,a}=0$. Expanding around the background and neglecting the metric fluctuations which are subdominant in this regime, one obtains after integration by parts 
\ba
S_{\rm two-deriv} = \int \d^4 x \sqrt{-g} \left[  \, -\frac{1}{2}(\p_{\mu} \delta \chi)^2 -\frac{1}{2} \left( \partial_{\nu} B^{a}_{\mu} \right)^2-\frac{ \lambda}{2f}\p_0 \bar \chi \epsilon^{ijk} B_{i}^{a}\p_j B^a_k-\frac{ \lambda}{2f} \delta\chi \p_0 (a \psi)\epsilon^{ijk}\p_i B^j_k \right] \nonumber 
\ea
The first two terms are the leading ones at high energies/momenta since for the last two terms, one of the derivatives acts on a background quantity $\bar \chi$, $a \psi$, and so are automatically subdominant in this regime. The first two terms are just the usual kinetic term for the axion, combined with the usual Lorenz-gauge kinetic term for a set of 3 $U(1)$ fields. This guarantees that for both the gauge field and axion fluctuations
\be
\lim_{k/a \gg m_g \gg H} \, c_s^2(k) \rightarrow 1 \, ,
\ee
where $c_s(k)$ is the speed of sound for a mode with comoving momenta $k$. 
Although it is possible that $c_s^2(k)$ can become negative for one of the modes in the region $m_g^2\gg k^2/a^2$, this does not indicate a gradient instability since the large mass term is sufficient to maintain stability of the solutions. It is only when $k^2/a^2$ dominates over $m_g^2$ that a negative $c_s^2$ signals an instability, and since we find $c_s^2(k) \approx 1$ in this region we see that there is no instability.  \\

Putting all of these results together, the absence of a gradient instability, the absence of any tachyonic instability as shown above, and the absence of ghosts which follows from working with a consistent theory at the outset, we can conclude that the SRSH limit of the chromo-natural inflationary cosmological solutions that we are considering is completely stable under small perturbations.


\section{Loop Corrections and Technical Naturalness}

In the previous section we integrated out the gauge fields in the SRHS limit only at tree level to determine $S_{\chi}$ and neglected the quantum fluctuations from the gauge field. The validity of this procedure requires that the masses of the all the fluctuations of the gauge field are of order $m_g \gg H$. If this is the case then decoupling tells us that we can neglect the gauge fields, even at the quantum level, and account for the effects of the gauge fields by local operators for $\chi$ in the Wilsonian action $S_W$. \\

The effective single field action $S_{\chi}$ is the tree level contribution to the Wilsonian action $S_W=S_{\chi}+ \Delta S_{\chi}$ obtained on integrating out the gauge fields\footnote{In this section we discuss loop corrections in the {\it in-out} path integral. Although technically we should perform this in the {\it in-in} path integral, the {\it in-in} corrections typically vanish faster (usually exponentially $e^{-c m_g^2/H^2}$ rather than power law $H^2/m_g^2$) than the {\it in-out} corrections in the limit $m_g \rightarrow \infty$. This is a simple consequence of the fact that particle creation of the heavy modes is exponentially suppressed in $H/m_g$ due to energy conservation. As long as the heavy particle effectively remains in vacuum, all of the quantum corrections are captured by the Wilsonian action derived from the {\it in-out} path integral. This Wilsonian action may then be quantized for the light fields in an {\it in-in} manner.}. At the quantum level we can state this as
\be
e^{i S _{\chi}+ i \Delta S_{\chi}} = \int D A_{\mu}^a \int D \mu_a \, J \, e^{i S_{\rm chromo} + i \int \d^4 x \, \mu_a f^{abi} A_{i}^b} \, ,
\ee
where $J$ is the Fadeev-Popov determinant and $\mu_a$ imposes the gauge condition. First let us consider the extreme SRHS limit to check that it is consistent at the quantum level. Since we are in the gauge $A_{i a}-A_{i a}=0$ we find that in the SRHS limit the Jacobian $J$ is equivalent to that coming from a local field redefinition and is simply unity $J=1$ in dimensional regularization. The gauge fixing terms may the discarded provided we remember to only integrate over the symmetric components of ${A}_i^a$ and set the antisymmetric ones to zero. We may then utilize the same locally inertial frame trick to simplify the form of the action and then covariantizing the result. The path integral over $\tilde A_0^a$ is Gaussian and may be easily performed, again giving rise to a measure contribution which vanishes in dimensional regularization. The final path integral over $\tilde A_i^a$ is less trivial. The relevant part is
\ba
&& e^{ i \Delta S_{\chi} } = e^{-i \int \d^4 x \sqrt{-g} \frac{1}{4 \Lambda^4} (\partial \chi)^4} \times \\ \nonumber
&& \int_{\rm sym} D {\tilde A}_{\mu}^a \exp{i \int \d^4 x \sqrt{-g}\left[ -\frac{1}{4}  (\tilde A_{i}^a \tilde A^{i}_a)^2+\frac{1}{4} (\tilde A_{i}^a \tilde A_{j}^a) (\tilde A^{i}_b \tilde A^{j}_b)  + \frac{\hat \lambda}{6f } \sqrt{- (\partial \chi)^2} \epsilon^{ij k} f^{abc} \tilde A_{i}^a \tilde A_{j}^b \tilde A_{k}^c \right]}
\ea
This path integral may be easily performed as a loop expansion around the saddle point given by the tree level solution
\be
\tilde {A}_i^a= \delta_i^a \frac{\hat \lambda}{2f} \sqrt{- (\partial \chi)^2}
\ee
The validity of this procedure is justified by the results of the previous section that show that the `mass-squareds' for the gauge field fluctuations are positive, and hence the path integral may be Wick rotated into a convergent one. However due to the absence of any derivative terms in this expression, the loop integrals around the saddle points again all vanish in dimensional regularization. This is a consequence of the fact that because the path integral is only over an algebraic function of $\tilde A_{\mu}^a$, we can always perform a field redefinition to make it gaussian (this may be demonstrated order by order), and in dimensional regularization the Jacobians from field redefinitions are unity. These arguments seem to suggest that the SRHS limit is fully consistent at the quantum level. \\

To some extent however, the previous argument is too quick. A slightly more careful consideration requires us to first regularize at finite $m_g$ and then take the SRHS limit of the regularized expression. In this case dimensional regularization gives us a contribution that survives\footnote{The fact that these two operations do not commute is a consequence of the adhoc way we discard divergences in dimensional regularization}. To one-loop order the quantum corrections coming from the fluctuations around the tree level solution can be schematically estimated using the Coleman-Weinberg potential (this is of course only the lowest order contribution in an infinite derivative expansion in $\partial/m_g$ which arises already at one-loop order)
\be
\Delta S_{\chi} \approx \int \d^4 x -C \,  \ln (m_g^2/\Box) m_g^4 + \sum_{m} a_m \left( \frac{\partial}{m_g} \right)^{2m} \, m_g^4 \, ,
\ee
where $C$ is an order unity coefficient. However this action must be expressed in terms of $\chi$ alone. Using the fact that in the background we know that $m_g = \sqrt 2 g \psi = \sqrt 2 {\lambda}\sqrt{-(\partial \chi)^2}/(2f)$ 
\be
S_W \simeq \int \d^4 x \sqrt{-g} \left[ - \frac{1}{2} (\partial \chi)^2 +\frac{\lambda^4}{32 f^4} \left( \frac{1}{g^2} -8 C \ln (m_g^2/\Box) \right) \, \left(  \partial \chi \right)^4- \mu^4 \left( 1+ \cos\left( \frac{\chi}{f}\right) \right)  + \dots \right] \, .
\ee
However it is easy to recognize that this is nothing other than the one-loop renormalization of the gauge field coupling. In other words, the running of the Wilson action under renormalization group flow coming from the gauge fields amounts to nothing more than the running of the coupling, so by renormalization group methods we can infer that:
\be
S_W \simeq \int \d^4 x \sqrt{-g} \left[ \frac{\mpl^2}{2} R - \frac{1}{2} (\partial \chi)^2 +\frac{\lambda^4}{32 f^4 g_r(\Box)^2}  \left(  \partial \chi \right)^4- \mu^4 \left( 1+ \cos\left( \frac{\chi}{f}\right) \right)  \right] + \dots \, ,
\ee
where $g_r(\Box)$ is the exact running coupling. Consequently the low energy EFT is technically natural in the sense that it is stable under renormalization from the loops of the massive gauge fields. This is nothing other than gauge invariance and the renormalizability of Yang-Mills theories at work. \\

The Wilson action $S_W$ is itself a non-renormalizable low energy EFT from axion loops due to the existence of the dimension 8 irrelevant operator $(\partial \chi)^4/\Lambda^4$. However these loops are fundamentally less dangerous since the axion is light (relative to the gauge bosons). Furthermore there is no difficulty making sense of $S_W$  as a low energy EFT. At first sight we might think that the cutoff for the low energy EFT should be taken to be $\Lambda$ as this is the scale of the suppression and the scale at which perturbative unitarity would be violated if this effective theory was used to calculate axion scattering in Minkowski space-time. However we know that this cannot be the case, the real cutoff must be determined by $m_g$ since that is the scale of the massive gauge fields we have integrated out. We can understand why this is the case from the low energy EFT as follows, considering just the action for the axion
\be
S_{\rm axion} = \int \d^4 x \sqrt{-g} \left[ -\frac{1}{2} (\partial \chi)^2 + \frac{1}{4 \Lambda^4} (\partial \chi)^4 - \mu^4 \left( 1+ \cos\left( \frac{\chi}{f}\right) \right) \right] \, ,
\ee
and expanding around the background $\chi= \bar \chi + \delta \chi$ the action for the fluctuations is very schematically (keeping track of only the time derivatives)
\be
S_{\rm axion} \simeq \int \d^4 x \sqrt{-g} \left[ -\frac{1}{2} \left(1 + 3 \frac{\dot {\bar \chi}^2}{\Lambda^4} \right)(\partial \delta \chi)^2+ \frac{\dot{\bar \chi}}{\Lambda^4} (\partial \delta \chi)^3 + \frac{1}{4 \Lambda^4} (\partial \delta \chi)^4 - \mu^4 \left( 1+ \cos\left( \frac{\chi}{f}\right) \right) \right] \, .
\ee
Working in the limit ${{\dot {\bar \chi}}^2} \gg {\Lambda^4}  $ and canonically normalizing the fluctuations 
\be
\delta \chi = v \Lambda^2/{\sqrt 3\dot {\bar \chi}}
\ee
we find that
\be
S_{\rm axion} \simeq \int \d^4 x \sqrt{-g} \left[ -\frac{1}{2} (\partial v)^2 +\frac{\sqrt 2 g}{6 \sqrt{3} m_g^3}  (\partial v)^3 + \frac{g^2}{18 m_g^4} (\partial v)^4 - \mu^4 \left( 1+ \cos\left( \frac{\chi}{f}\right) \right) \right] \, .
\ee
This shows that the maximum possible cutoff of the EFT is the scale of perturbative unitarity violation which is $m_g/g^{1/3}$ (since we require $g<1$) and not $\Lambda$. Including higher loops into this argument eventually brings it down to the scale $m_g$. This is completely consistent with the known UV completion of this low energy effective theory which in this case is simply $S_{\rm chromo}$\footnote{By UV completion here we mean the partial UV completion of the single field effective theory. The chromo-natural Lagrangian must itself be UV completed at a higher energy scale $f/\lambda$.}. This is an explicit example of when the cutoff of the low energy EFT is determined by the background about which we are expanding. It is also an explicit example of the Vainshtein effect, whereby dressing the kinetic term for fluctuations raises the effective cutoff of the theory\footnote{For a discussion of how this works in the the simplest Galileon model see \cite{Nicolis:2004qq}}. \\

Finally let us comment on the general structure of the loop corrections from the heavy fields. Since the true cutoff for the low energy axion EFT is $m_g$ - which may be expressed covariantly in terms of $\chi$ as $m^2_g= -{\lambda^2}(\partial \chi)^2/f^2$ - the structure of the Wilson action to all loop order is expected to be of the schematic form
\ba
\label{loop1}
S_W &\simeq& \int \d^4 x \sqrt{-g} \left[ \frac{\mpl^2}{2} R - \frac{1}{2} (\partial \chi)^2 +\frac{\lambda^4}{32 f^4 g^2}  \left(  \partial \chi \right)^4- \mu^4 \left( 1+ \cos\left( \frac{\chi}{f}\right) \right)  \right]   \\
 &&+ \sum_{n,m,p} c_{nmp} \,  \left(\frac{\lambda^2}{f^2}(\partial \chi)^2 \right)^2 \left( \frac{\partial}{\sqrt{-\frac{\lambda^2}{f^2}(\partial \chi)^2}} \right)^{2 m}  \left( \frac{f^2 \partial \chi}{\lambda^2 (\partial \chi)^2}\right)^n \left( \frac{R_{\mu \nu \rho \sigma}}{\frac{\lambda^2}{f^2}(\partial \chi)^2 } \right)^p  \nonumber \, .
\ea
This is further justified by recognizing that we can rearrange this expression as\footnote{The map between these two expressions is highly nontrivial. Strictly speaking in (\ref{loop1})  we were working implicitly with $\dot{\chi}^2 \gg \Lambda^4$ but to rewrite it as (\ref{loop2}) we must allowing for the regime $\dot{\chi}^2 < \Lambda^4$ by returning to the original Lagrangian. However, all that is really relevant about (\ref{loop1}) is that the dimensionful scale entering in the expression for the loop corrections is $f/\lambda$, which is consistent with the cutoff of the chromo-natural theory (strictly speaking this is the scale for perturbative unitarity violation for scattering of axions and gauge bosons in Minkowski space-time). In the end the Yang-Mills coupling $g$ plays a less significant role since in the limit $f/\lambda \rightarrow \infty$ for fixed $\lambda$ the chromo-natural theory $S_{\rm chromo}$ is renormalizable. Thus $g$ corrections alone do not induce non-renormalizable operators.}
\ba
\label{loop2}
S_W &\simeq& \int \d^4 x \sqrt{-g} \left[ \frac{\mpl^2}{2} R - \frac{1}{2} (\partial \chi)^2 +\frac{\lambda^4}{32 f^4 g^2}  \left(  \partial \chi \right)^4- \mu^4 \left( 1+ \cos\left( \frac{\chi}{f}\right) \right)  \right] \nonumber \\
 &&+ \sum_{n,m,p} d_{nmp} \, \frac{f^4}{\lambda^4}   \left( \frac{\partial}{f/\lambda} \right)^{m}   \left( \frac{ \partial \chi }{f^2/\lambda^2 } \right)^n  \left( \frac{R_{\mu \nu \rho \sigma}}{f^2/\lambda^2 } \right)^p\, ,
\ea
in which we see that the chromo-natural scale $f/\lambda$ enters. The chromo-natural scale $f/\lambda$ is the scale of perturbative unitarity violation for axion-gauge field scattering in Minkowski space-time for the original chromo-natural Lagrangian. That this is true can be seen by direct inspection of the action $S_{\rm chromo}$ by noting that the axion-gauge field coupling is a dimension 5 operator suppressed by the scale $f/\lambda$.

\section{Comparison with Gelaton Mechanism}
\label{gelaton}
Chromo-natural inflation can be understood as an example of the {\it gelaton} mechanism \cite{Tolley:2009fg} (see also \cite{Achucarro:2010jv, Achucarro:2010da, Achucarro:2012sm, Achucarro:2012yr, Burgess:2012dz}). 
The idea of the gelaton is that a heavy field (in this case played by the gauge fields) has a significant effect on the dynamics of the light field (in this case the axion) because even though the mass of the heavy field is large $m_g \gg H$, the heavy field is `reasonably' strongly coupled to the light field. By `reasonably' strong we mean large enough that the coupling terms have an $O(1)$ correction to the action, more precisely to the kinetic term of the action, without the coupling being so large that the quantum theory goes out of control. This large coupling forces the heavy field to gel to the dynamics of the light field. In turn the dynamics of the light field is strongly influenced by the energy density of the field. The net result is that the dynamics of the light field is described by a low energy EFT with a non-minimal kinetic term. 

A simple example of the gelaton mechanism in the case of a two scalar field theory is given by (see \cite{Tolley:2009fg} for similar examples)
\be
S_{\rm gelaton}= \int \d^4 x \left[ -\frac{1}{2} \left( 1 + \frac{g \Phi^2}{2 \Lambda^2} \right)(\partial \chi)^2-V(\chi) - \frac{1}{2} (\partial \Phi)^2-\frac{g^2}{8} \Phi^4 \right]
\ee
Varying with respect to the heavy field (the gelaton) $\Phi$ and assuming that the heavy field fluctuations are sufficiently massive that $\Phi$ sits at the bottom of its effective potential
\be
V_{\rm eff}(\Phi) =  \frac{g \Phi^2}{4 \Lambda^2} (\partial \chi)^2 +\frac{g^2}{8} \Phi^4  \,  ,
\ee
we infer the equation of motion for the heavy field
\be
2 V_{\rm eff}'(\Phi)=\frac{g \Phi}{\Lambda^2}(\partial \chi)^2+g^2 \Phi^3 \approx 0 \, ,
\ee
whose nontrivial solution is
\be
\Phi^2 \approx -\frac{1}{g \Lambda^2} (\partial \chi)^2 \, .
\ee
Substituting back into the action and again neglecting the kinetic/gradient term for the heavy field we find
\be
S_{\rm gelaton}= \int \d^4 x  \left[  X+ \frac{1}{\Lambda^2} X^2-V(\chi) \right] 
\ee
where $X=-(1/2)(\partial \chi)^2$ which is precisely the form of the effective field theory for chromo-natural inflation. The validity of these assumptions relies on the mass of the gelaton being large. In this model the mass is given by $m_{\Phi}^2=V''_{\rm eff}(\Phi)\sim {g}| (\partial \chi)^2|/{\Lambda^2}$. The limit in which the gelaton decouples corresponds to the scaling limit $g \rightarrow \infty$. This is the precise analogue of the SRHS limit for chromo-natural inflation. Again as in this case, the mathematical scaling limit does not necessarily imply $g$ is larger than unity, but rather encodes the physical requirement that
\be
g \gg \frac{\Lambda^2 H^2}{ |(\partial \chi)^2|} = \frac{H^2}{\Lambda^2}  \frac{\Lambda^4}{ |(\partial \chi)^2|} 
\ee
which may be easily satisfied since we can have $|(\partial \chi)^2|\gg \Lambda^4$ even if $H \sim \Lambda$.

\section{Power Spectrum, Bispectrum}
Having estabilished the stability of the model and its equivalence in the SRHS regime with a generic single field $P(X,\chi)$, we now proceed to computing its predictions for the primordial power spectrum and bispectrum. Using the language of \cite{Seery:2005wm,Chen:2006nt} for a Lagrangian 
\be
S=\int \d^{4}x\sqrt{-g}\left[\frac{\mpl^{2}}{2}R+P(X,\chi)\right],
\ee
the Einstein equations can be put in the following form
\be
3\mpl^{2} H^{2}=\rho,
\ee
\be
\dot{\rho}=-3H\left(\rho+P\right),
\ee
where 
\be
\rho=2 XP_{,X}-P
\ee
and $P_{,X}$ denotes a derivative w.r.t. X (here $X\equiv-\left(\partial\chi\right)^{2}/2$). We can define a ``speed of sound'' $c_{s}$
\be
c_{s}^{2}=\frac{P_{,X}}{P_{,X}+2XP_{,XX}}
\ee
and the slow-roll parameter is
\be
\epsilon\equiv-\frac{\dot{H}}{H^2}=\frac{XP_{,X}}{H^{2}\mpl^{2}}.
\ee
The power spectrum of the primordial fluctuation is \cite{Garriga:1999vw}
\be
P_{k}^{\zeta}=\frac{1}{36\pi^{2}\mpl^{4}}\frac{\rho^{2}}{c_{s}\left(\rho+P\right)}=\frac{1}{8\pi^{2}\mpl^{2}}\frac{H^{2}}{c_{s}\epsilon},
\ee
which is evaluated at horizon exit $c_{s}K=aH$.\\
The tensor perturbation spectrum is
\be
P_{k}^{h}=\frac{2}{3\pi^{2}}\frac{\rho}{M_{Pl}^{4}}.
\ee
The bispectrum for this class of models was computed in \cite{Chen:2006nt}; we report the result to leading order in slow-roll
\be
\label{bspt}
\langle\zeta(\vec{k}_{1})\zeta(\vec{k}_{2})\zeta(\vec{k}_{3})  \rangle=(2\pi)^{7}\delta^{(3)}\left(\vec{k}_{1}+\vec{k}_{2}+\vec{k}_{3}\right)\left(P_{k}^{\zeta}\right)^{2}\frac{\mathcal{A}_{\Omega}+\mathcal{A}_{c}}{\Pi_{i}k_{i}^{3}},
\ee
where
\be
\mathcal{A}_{\Omega}\simeq \left(\frac{1}{c_{s}^{2}}-1-\frac{2\Omega}{\Sigma}\right)\frac{3k_{1}^{2}k_{2}^{2}k_{3}^{2}}{2K^{3}},
\ee
\be
\mathcal{A}_{c}\simeq\left(\frac{1}{c_{s}^{2}}-1\right)\left(-\frac{1}{K}\sum_{i>j}k_{i}^{2}k_{j}^{2}+\frac{1}{2K^{2}}\sum_{i\neq j}k_{i}^{2}k_{j}^{3}+\frac{1}{8}\sum_{i}k_{i}^{3}\right),
\ee
with $K\equiv k_{1}+k_{2}+k_{3}$ and $\Sigma$ and $\Omega$ defined as
\be
\Sigma\equiv XP_{,X}+2X^{2}P_{,XX}
\ee
\be
\Omega\equiv X^{2}P_{,XX}+\frac{2}{3}X^{3}P_{,XXX}.
\ee
In our specific model
\be
P(X,\chi)=X+\frac{X^{2}}{\Lambda^{4}}-V\left(\chi\right),
\ee
where $\Lambda^{4}=8f^{4}g^{2}/\lambda^4$ and $V\left(\chi\right)=\mu^{4}\left(1+\cos(\chi/f)\right)$. From the definitions above we have
\be
\rho=X+\frac{3X^{2}}{\Lambda^{4}}+V
\ee
and therefore the slow-roll condition $\epsilon\simeq(\rho+P)/\rho\ll 1$ implies $\rho\simeq V$, i.e. that the total energy is dominated by the potential of $\chi$ ($\rho\simeq V$). From Einstein equations we also have
\be
\dot{\rho}=-3H\left(2X+\frac{4X^{2}}{\Lambda^{4}}\right)=V_{,\chi}\dot{\chi},
\ee
which is equivalent to 
\be
3H\dot{\chi}\left(1+\frac{\dot{\chi}^{2}}{\Lambda^{4}}\right)-\frac{\mu^{4}}{f}\sin\left(\frac{\chi}{f}\right)=0.
\ee
from Sec.(3). 
We can use this expression to determine $\dot{\chi}$ as a function of $\chi$ which, assuming the cubic term dominates so that the enhances damping is in effect simplifies to
\be
\dot{\chi}\simeq \frac{2fg^{2/3}\mu^{4/3}\sin^{1/3}(\chi/f)}{3^{1/3}H^{1/3}\lambda^{4/3}}.
\ee
This result will be used in computing the power spectrum, spectral index and number of e-foldings. Before we do so, note that the requirement that the damping is enhanced
\be
X=\frac{\dot{\chi}^{2}}{2}=\frac{2f^{2}g^{2}\psi^{2}}{\lambda^{2}}\gg \Lambda^{4},
\ee
is equivalent to the requirement that $\lambda\psi/f\gg 1$. With this in mind, we obtain the following results for chromo natural inflation in SRHD regime
\be
c_{s}^{2}=\frac{1}{3},\quad\quad\quad\quad \epsilon=\frac{2X^{2}}{H^{2}\mpl^{2}\Lambda^{4}}.
\ee
The primordial power spectrum reads
\be
P_{k}^{\zeta}=\frac{\sqrt{3}}{144\pi^{2}}\frac{\Lambda^{4}V^{2}}{\mpl^{4}X^{2}}=\frac{3^{7/6}\mu^{16/3}\lambda^{4/3}\left(1+\cos(\chi/f)\right)^{8/3}}{72\pi^{2}\mpl^{16/3}g^{2/3}\left(\sin(\chi/f)\right)^{4/3}}, \label{powerspectrum}
\ee
from which we can compute the spectral tilt $n_{s}$ 
\be
n_{s}-1=\frac{d\ln P_{k}^{\zeta}}{d\ln k}=\frac{1}{P_{k}^{\zeta}}\frac{\dot{\chi}}{H}\frac{d P_{k}^{\zeta}}{d\chi}=\left(\frac{2\times  3^{1/3}\mpl^{4/3}g^{2/3}}{\mu^{4/3}\lambda^{4/3}}\right)\frac{8\cos^{2}(\chi/2f)\left(\cos(\chi/f)-2\right)}{3\sin^{2/3}(\chi/f)\left(1+\cos(\chi/f)\right)^{5/3}}. \label{spectralindex}
\ee
The power spectrum from tensor modes is
\be
P_{k}^{h}=\frac{2}{3\pi^{2}}\left(\frac{\mu}{M_{Pl}}\right)^{4}\left(1+\cos\left(\chi/f\right)\right).
\ee
The number of e-foldings is given by
\be
N_e=\int_{t_{i}}^{t_{f}} H dt=\int_{\chi_{i}}^{\chi_{f}} \frac{H}{\dot{\chi}}d\chi=\int_{\chi_{i}}^{\chi_{f}}\left(\frac{\mu^{4}\lambda^{4}}{24f^{3}g^{2}\mpl^{4}}\right)^{1/3}\frac{\left(1+\cos(\chi/f)\right)^{2/3}}{\sin^{1/3}(\chi/f)}d\chi=\left(\frac{9}{4}\frac{\mu^{4}\lambda^{4}}{g^{2}\mpl^{4}}\right)^{1/3}
\ee
where we integrated in the range $\chi/f=x\in [0,\pi]$. \\
From here we can express the spectral index in terms of the number of e-foldings as
\be
n_{s}-1=-\frac{\mathcal{C}}{N_e}.
\ee
where the coefficient $\mathcal{C}$ is a positive function of $\chi/f$ that we can read read off Eq.~(\ref{spectralindex}). We know from observations that $n_{s}=0.963\pm 0.012$ \cite{Larson:2010gs}, so if we take the central value for it, we have
\be
N_e\simeq 27\times\mathcal{C}.
\ee
Similarly, the power spectrum can be put in the form
\be
P_{k}^{\zeta}=\left(\frac{\mu}{\mpl}\right)^{4}\mathcal{B}N_e \, .
\ee
with a (positive) coefficient $\mathcal{B}$ defined from Eq.~(\ref{powerspectrum}). 
To get an idea, some typical values for $\mathcal{B}$ and $\mathcal{C}$ are respectively $0.05$ and $4.4$, which comfortably allow for our model to produce predictions for the relevant observables that are within experimental bounds. $\mathcal{C}$ is characterized by a plateau in the interval $\chi/f\in[0.1,2.3]$ and it ranges between $\sim 4$ and $\sim 14$; the range of values of $\mathcal{B}$ in the same interval of $\chi/f$ is between $\sim 0.001$ and $\sim 0.7$. We  keep these quantities abstract as $\mathcal{B}$ and $\mathcal{C}$ in our results, instead of explicitating them, in order to underline the fact that there is further freedom in the model.\\ By setting $P_{k}^{\zeta}\simeq 2\times 10^{-9}$, we can obtain some conditions on our parameters
\be
\label{log}
\frac{\mu}{\mpl}\simeq\frac{3\times 10^{-3}}{\left(\mathcal{B}\mathcal{C}\right)^{1/4}},\quad\quad\quad\quad \frac{\lambda^{2}}{g}\simeq 1.6\times 10^{7}\mathcal{B}^{1/2}\mathcal{C}^{2}.
\ee
The main inequalities that need to be satisfied are
\be
\epsilon_{2}\gg \epsilon_{1}\quad\quad\rightarrow g^{2}\psi^{2}\gg H^{2},
\ee
\be
g^{2}\psi^{4}\ll \mu^{4}\simeq H^{2}\mpl^{2}\quad\rightarrow g^{2}\psi^{2}\ll H^{2}\left(\frac{\mpl}{\psi}\right)^{2},
\ee
\be
\label{flpsi}
\frac{\lambda\psi}{f}\gg 1 \, .
\ee
One condition we can infer from the above is the following
\be
g\gg \frac{2\times 10^{-5}}{\mathcal{B}^{1/2}}.
\ee
Using the definition of $\psi_{min}$ together with the second of Eqs.~(\ref{log}), we get
\be
\epsilon_{2}\simeq \frac{5\times 10^{-2}}{\mathcal{C}},\quad\quad\quad \epsilon_{1}=\frac{10^{-6}}{g\mathcal{B}^{1/2}\mathcal{C}}.
\ee
From the conditions in Eq.~(\ref{flpsi}) on $f$ we have
\be
 \frac{f}{\mpl}\ll 4 \mathcal{C}^{1/2}.
\ee
which is not only straightforward to satisfy, it is desirable for any realistic model. 
The tensor-to-scalar ratio is given by
\be
r\equiv\frac{P^{h}_{k}}{P_{k}^{\zeta}}=\left(\frac{2}{3\pi^{2}}\right)\frac{1+\cos\left(\chi/f\right)}{\mathcal{B}N_{e}}.
\ee
Typical values of $r$ for $\mathcal{B}=0.05$ and $\mathcal{C}=4.4$ are in the interval $[0.01,0.03]$, for $n_{s}$ varying in its range $[0.951,0.975]$. 
Finally let us check the value of the slow-roll parameter $\epsilon_{\chi}$
\be
\epsilon_{\chi} \simeq \frac{10^{-3} f^2}{\mathcal{C}^2 \mpl^2} \ll 1
\ee 
Putting this all together we see that there is no difficulty satisfying all of the appropriate slow-roll conditions and fitting the observed power spectrum parameters with this model and obtaining sufficient number of e-folds of inflation. \\

\noindent For the bispectrum, we parametrize the amplitude with $f_{NL}$ defined as in 
\be
\langle\zeta(\vec{k}_{1})\zeta(\vec{k}_{2})\zeta(\vec{k}_{3})  \rangle=(2\pi)^{7}\delta^{(3)}\left(\vec{k}_{1}+\vec{k}_{2}+\vec{k}_{3}\right)\left(-\frac{3}{10}f_{NL}\left(P_{k}^{\zeta}\right)^{2}\right)\frac{\sum_{i}k_{i}^{3}}{\Pi_{i}k_{i}^{3}},
\ee
and insert
\be
\Sigma=\frac{6X^{2}}{\Lambda^{4}},\quad\quad\quad\quad\Omega=\frac{2X^{2}}{\Lambda^{4}}
\ee
in Eq.~(\ref{bspt}). We obtain two contributions to $f_{NL}$
\be
f_{NL}^{\Omega}\simeq -0.08 ,\quad\quad\quad\quad f_{NL}^{c}\simeq 0.65,
\ee
both in the equilateral configuration (the bispectrum (\ref{bspt}) peaks in the equilateral limit $k_{1}\simeq k_{2}\simeq k_{3}$). These values are well whithin experimental limits considering the available observational bounds $-214<f_{NL}^{equil}<266$ ($95\% CL$) \cite{Komatsu:2010fb}.\\

\section{Beyond the SRHS Limit}

\label{beyond}

In this section we shall determine the correct single field EFT is away from the SRHS limit, assuming that there exists a range of slow-roll parameters for which the gauge fields continue to be sufficiently massive that they can be neglected during inflation. Unlike the SRHS limit we shall not provide a rigorous derivation of this single field EFT but rather utilize what we have learned from the preceding analysis. 

As long as we are in a slow-roll regime in which the axion potential dominates in the Friedman equation the effective equations for $\chi$ and $\psi$ are 
\ba
&&3 H \dot{\chi} - \frac{\mu^4}{f} \sin \left( \frac{\chi}{f}\right)=-3 g \frac{\lambda}{f} H \psi^3  \, ,\\ 
&& 2 H^2 \psi + 2 g^2 \psi^3 = g \frac{\lambda}{f} \psi^2 \dot{\chi} \, .
\ea
In writing these equations we have neglected terms with $\ddot{\chi}$ and $\dot{\psi}$ and $\ddot{\psi}$. The reason for neglecting $\dot{\psi}$ is that the will be related to $\ddot{\chi}$ and hence will be slow-roll suppressed. Again in the slow-roll approximation we have
\be
H^2 \simeq \frac{1}{3 \mpl^2} \mu^4 \left( 1 + \cos \left( \frac{\chi}{f}\right) \right) \, .
\ee
To derive the single field EFT we rewrite these two equations as a single equation for the axion. Solving for $\psi$ and choosing the root that is continuous with the SRHS limit we have
\be
\psi = \frac{1}{4 f g} \left(\lambda \dot{\chi} + \sqrt{\lambda^2 \dot{\chi}^2-\frac{16}{3 \mpl^2} f^2 \mu^4  \left( 1 + \cos \left( \frac{\chi}{f}\right) \right)}  \right) \, .
\ee
Substituting this back into the previous equation gives
\be
3 H \left[ 1 + \frac{\lambda^4}{32 f^4 g^2} \dot{\chi}^2 \left( 1 + \sqrt{1-\frac{8}{3 \mpl^2 \lambda^2 \dot \chi^2} f^2 \mu^4  \left( 1 + \cos \left( \frac{\chi}{f}\right) \right)} \right)^3 \right] \dot{\chi} - \frac{\mu^4}{f} \sin \left( \frac{\chi}{f}\right) =0 \, .
\ee
We now ask what single field EFT gives rise to this slow-roll equation. This is simple to deduce, as before it is a $P(X,\chi)$ model since the slow-roll equation for any $P(X,\chi) $ model is given by
\be
3 H \left( \frac{\partial }{\partial X}P(X,\chi) \right) \dot{\chi}= \frac{\partial}{\partial \chi} P \, .
\ee
Thus we have to solve
\be
\frac{ \frac{\partial }{\partial X}P(X,\chi)  }{ \frac{\partial}{\partial \chi} P(X,\chi)} = \frac{\left[ 1 + \frac{\lambda^4}{32 f^4 g^2} X \left( 1 + \sqrt{1-\frac{8}{3 \mpl^2 \lambda^2 X} f^2 \mu^4  \left( 1 + \cos \left( \frac{\chi}{f}\right) \right)} \right)^3 \right] }{ \frac{\mu^4}{f} \sin \left( \frac{\chi}{f}\right) } \, .
\ee
where $X = - (\partial \chi)^2/2$. 
Thus the full single field EFT can be obtained by integrating this equation. The final result is cumbersome, but expanding gives terms if the form
\be
P(X,\chi) = X+\frac{1}{ \Lambda^4 } X^2 -\mu^4  \left( 1 + \cos \left( \frac{\chi}{f}\right) \right) + \frac{1}{ \Lambda^4 } X^2  \left( \frac{f^2 \mu^4}{\mpl^2 \lambda^2 X} \right)^n \text{corrections} \, .
\ee
Which is to say that the corrections are of the form $\frac{1}{ \Lambda^4 } X^2 \left( \frac{H^2}{m_g^2}\right)^n$ as we would expect. 

Since this single field EFT is sufficient to reproduce the correct slow-roll equations of motion for the background, it will be sufficient to use this to calculate the power spectrum and non-Gaussianities as long as we can assume the gauge fields decouple. We see that the form of the $P(X,\chi)$ mode is such that we require
\be
\frac{16}{3 \mpl^2 \lambda^2 \dot \chi^2} f^2 \mu^4  \left( 1 + \cos \left( \frac{\chi}{f}\right) \right)=\frac{16 H^2}{ \lambda^2 \dot \chi^2} f^2  \lesssim 1
\ee
otherwise the square root becomes ill-defined. This is essentially the statement that 
\be
m_g^2 \gtrsim 8  H^2  \text{ or } \epsilon_2 \gtrsim 4 \epsilon_1 \, .
\ee
Thus based on these observations the single field EFT for chromo-natural inflation is only capable of describing slow-roll inflation if $\epsilon_2 \gtrsim 4 \epsilon_1$. As long as this condition is satisfied then the above single field EFT is appropriate. But in turn this condition is the necessary condition for the gauge fields to decouple during inflation. If $\epsilon_2 < 4 \epsilon_1$ then a single field EFT is not appropriate and it will be necessary to solve the full dynamics of the axion and gauge fields. This is beyond the scope of this paper.

\section{Conclusions}

Chromo-natural inflation is a compelling theory for inflation which has very recently been proposed \cite{Adshead:2012kp}. The crucial ingredients of the model, are an axion coupled to an $SU(2)$ gauge field in a manner which typically arises in UV complete models. The axion-gauge field coupling, creates an additional damping or friction term in the slow-roll equation for the axion, making it easier to satisfy the slow-roll requirements of inflation, with coupling constants set to realistic values. \\

In this paper we have studied a slow-roll scaling limit of this inflationary model. The crucial result is that in this regime the gauge fields are sufficiently massive that they can be safely integrated out whilst, at the same time, the non-trivial modification of the axion dynamics survives the integration process in the form of a modified kinetic term for the resulting single-field effective field theory. This is a clear-cut realization of the {\it gelaton} mechanism \cite{Tolley:2009fg}. We have demonstrated both the classical and quantum consistency of the resulting low energy effective quantum field theory: the masses of all the gauge field fluctuations are positive and well above the Hubble scale $H$. A scaling limit for the gauge fields in then justified, their dynamics being recast as local operators in the axion Wilsonian action. In \textit{Section 5} we have seen how the resulting effective field theory is stable under renormalization from the loops of gauge fields, and have estimated the typical form of loop corrections.\\ 

The single-field effective action we obtained nicely lends itself to a straightforward estimate of the principal relevant observables for inflation predicted by this model. Our results are all well within experimental values. The COBE normalization for the power spectrum, the WMAP value for the spectral tilt and the lower bound on the number of e-folds are all used to fix the model parameters. Interestingly some freedom still survives as detailed in \textit{Section 7}. In particular there is sufficient freedom to have the axion decay constant at a natural scale.
One naturally wonders what happens beyond the scaling limit of the theory. In the last section we have seen how, well away from the SRHS regime, a study of the full dynamics of the axion and gauge fields is required. From our analysis, it is not clear that away from SRHS the theory will prove to behave as nicely. An interesting on going investigation on \textit{gauge-flation} \cite{Maleknejad:2011jw,Maleknejad:2011sq}, which can be seen as a different limit (in gauge-flation  one integrates out the axion) of chromo-natural inflation, is to be found  in \cite{Maleknejad:2011jr,Maleknejad:2012as,SheikhJabbari:2012qf,Noorbala:2012fh}. 
A study of the stability for the full chromo-natural model would be required (including gradient instability), and scale invariance itself is not guaranteed. We leave these interesting questions to future work. 

\acknowledgments

We would like to thank Peter Adshead and Mark Wyman for useful discussions. ED is very happy to thank Marco Peloso for many fruitful discussions. ED would very much like to thank the CWRU Physics Department for warm hospitality during several stages of this work. AJT is supported by DOE grant DE-FG02-12ER41810. ED is partially supported by DOE grant DE-FG02-94ER-40823.

\end{document}